\documentclass[twocolumn,prl,aps]{revtex4}
\usepackage{amsmath}   
\usepackage{amssymb}
\usepackage{amsfonts}
\usepackage{mathrsfs}
\usepackage{graphicx}

\usepackage{siunitx} 

\begin{document}

\title
{Achieving Balance of Valley Occupancy in Narrow AlAs Quantum Wells}

\author{A.~R.~Khisameeva$^{a,b}$, A.~V.~Shchepetilnikov$^{a}$, V.~M.~Muravev$^{a}$ \footnote{Corresponding author muravev@issp.ac.ru},  S.~I.~Gubarev$^{a}$, D.~D.~Frolov$^{a}$, Yu.~A.~Nefyodov$^{a}$, I.~V.~Kukushkin$^{a,c}$, C.~Reichl$^{d}$, W.~Dietsche$^{d}$, W.~Wegscheider$^{d}$}
\affiliation{$^a$ Institute of Solid State Physics, RAS, Chernogolovka, 142432 Russia\\ 
$^b$ Moscow Institute of Physics and Technology, Dolgoprudny, 141700 Russia \\
$~c$ National Research University Higher School of Economics, Laboratory for Condensed Matter Physics, Moscow, 101000 Russia \\
$^d$ Solid State Physics Laboratory, ETH Zurich, Otto-Stern-Weg 1, 8093 Zurich, Switzerland \\}
\date{\today}

\date{\today}

\begin{abstract}
Terahertz photoconductivity of $100~\mu$m and $20~\mu$m Hall bars fabricated from narrow AlAs quantum wells (QWs) of different widths is investigated in this paper. The photoresponse is dominated by collective magnetoplasmon excitations within the body of the Hall structure. We observed a radical change of magnetoplasma spectrum measured precisely for AlAs QWs of width ranging from $4$~nm up to $15$~nm. We have shown that the observed behavior is a vivid manifestation of valley transition taking place in the two-dimensional electron system. Remarkably, we show that photoresponse for AlAs QWs of width $6$~nm features two resonances, indicating simultaneous occupation of strongly anisotropic $X_{x-y}$ valleys and isotropic $X_z$ valley in the QW plane. Our results pave the way to realizing valley-selective layered heterostructures, with potential application in valleytronics.

\end{abstract}

\maketitle

\section{INTRODUCTION}

In the past five decades, Si-based semiconductor electronics has gone through an impressive increase in development. However, further progress in silicon technology has revealed fundamental physical limitations. Thus, there is an active search for novel device concepts. Spintronic devices, for example, make use of the spin of electrons~\cite{Wolf:01, Zutic:04}. Valleytronics is a more recent development that harnesses electrons' valley degree of freedom instead of the conventional charge-based operations~\cite{Gunn, Shayegan:06, Shayegan:07, Grayson:08, Grayson:11, Grayson:12,  Zeng:12, Jones:13, Shayegan:18}. One of the first valleytronic device was the Gunn diode, whose operation relies on the intervalley transfer of electrons in the presence of a strong electric field~\cite{Gunn}.  

A perspective material for research and realization of valleytronic concepts is the AlAs two-dimensional electron system (2DES)~\cite{Shayegan:06}. Its valley degree of freedom stems from the three-fold degenerate valleys located at the X-points of the Brillouin zone along [100], [010], and [001] crystallographic directions. These valleys are traditionally referred to as $X_x$, $X_y$, and $X_z$, respectively. The corresponding anisotropic, elliptical Fermi contours are characterized by a heavy longitudinal mass $m_{\rm l} = 1.1 m_0$ and a light transverse mass $m_{\rm tr}= 0.2 m_0$, where $m_0$ is the electron mass in a vacuum. The valley occupancy in AlAs quantum wells (QWs) can be tuned via uniaxial strain~\cite{Gunawan:06} or confinement~\cite{Kesteren:89, Yamada:94, Gunawan:06-2, Khisameeva:18}.   

In case a QW is grown along the [001] direction, the symmetry of the system is reduced and the degeneracy of the valleys is lifted by both the confinement potential and the inevitable built-in strain. Note that confinement favors the occupation of the $X_z$ valley with electrons, as this valley is characterized by the largest mass in the growth direction. In contrast, the biaxial compression of the AlAs layer arising from the lattice constant mismatch lowers the energies of the two in-plane valleys. The interplay between these two factors establishes the actual distribution of electrons between the valleys. In the quantum wells of width $W > 5$~nm, the contribution from the biaxial strain is dominant and the in-plane $X_{x-y}$ valleys are occupied, whereas electrons confined in a narrow ($W < 5$~nm) AlAs QW occupy a single out-of-plane $X_z$ conduction-band valley~\cite{Kesteren:89, Yamada:94, Khisameeva:18}.

The AlAs quantum wells stand apart from all other members of two-dimensional electron system (2DES) family realized in semiconductor heterostructures because of the possibility to tune valley occupancy. Although, the $\Gamma - X$ energy spectrum transition in narrow AlAs QWs has been quantitatively established, systematic study of system behavior near this transition is missing. Especially important is the range of QW widths when both $X_z$ and $X_{x-y}$ valleys are occupied simultaneously. Standing at this critical region, one can switch between the valleys in a controllable way by external straining or electric field. Therefore this region is especially important for possible valleytronic device concepts.

\section{EXPERIMENTAL METHOD}

\begin{table}[h!]
 \begin{center}
    \caption{Characteristics of AlAs/AlGaAs heterostructures}
   \label{tab:table1}

   \renewcommand{\arraystretch}{1.2}%
    \begin{tabular*}{\columnwidth}{@{\extracolsep{\fill}}|c|c|c|}
     \hline
      \textbf{QW's width} & \textbf{Density (1.5~K)} & \textbf{Mobility (1.5 K)}\\
      $W$ (nm) & $n_{s}$ \, ($10^{11} \,\text{cm}^{-2}$) & $\mu$ ($10^{3} \,\text{cm}^{2}/\text{V} \cdot \text{s}$)\\
      \hline
      4 & 7.7 & 14 \\
      4.5 &  4.6  & 34 \\ 
      5 & 5.5 & 41 \\
      5.5 &  7.0  &  19  \\
      6 & 9.2 & 9 \\
      6.5 & 8.0 & 16 \\
      7.0 & 4.6 & 30 \\
      15  & 1.5 & 150 \\ 
      \hline
    \end{tabular*}
  \end{center}
\end{table}

The samples under study were fabricated from a AlAs/Al$_{x}$Ga$_{1-x}$As ($x=0.46$) heterostructures grown along the $[001]$ direction via molecular beam epitaxy. The heterostructures contained a QW hosting the 2DES of one of the widths $W = 4{.}0$, $4{.}5$, $5{.}0$, $5{.}5$, $6{.}0$, $6{.}5$, $7{.}0$~nm, and $15$~nm. Each QW is doped from both sides, and the structures contain two Si doping layers: upper and lower. The upper spacer thickness has been fixed at $35$~nm. The distance between lower doping and QW was $400$~nm for the wide QW sample ($W=15$~nm) and $175$~nm for the narrow QWs ($W=4-7$~nm). The electron density, $n_s$, and electron transport mobility, $\mu$, for each structure are listed in Table~\ref{tab:table1}. The $W=4 - 7$~nm QW structures have been illuminated by a green light emitting diode, whereas $W=15$~nm structure has been measured without illumination. The sample was patterned into a Hall bar geometry using standard photolithography. Two different types of Hall bar geometry were used during the experiments. The first set of measurements was performed on $L=100$~$\mu$m wide Hall bars with a distance between voltage probes of $1.0$~mm. The Hall bar body was oriented along the $[110]$ crystallographic direction. The second set of measurements was carried on $L=20$~$\mu$m wide Hall bars with the distance between voltage probes of $1.0$~mm. Three Hall bars were oriented along the $[010]$, $[100]$, and [110] directions (see Fig.~3). Contacts to the 2DES were made by soldering indium to the Hall bar contact pads. The magnetoplasma excitations were detected using a double lock-in technique~\cite{Vasiliadou:93, Khisameeva:18}. We measured the microwave-induced part of longitudinal resistance $\Delta R_{xx}$, which appears due to resonant microwave absorption. The magnetoresistance $R_{xx}$ was measured by driving a $1~\mu$A sinusoidal current at a frequency of $\sim 1$~kHz between source and drain contacts of the Hall bar. The differential $\Delta R_{xx}$ signal was extracted from modulation ($f_{\rm mod} = 31$~Hz) of  the microwave radiation incident on the Hall bar structure. The magnetoresistance was acquired under microwave radiation from a set of generators, which covered the frequency range from $20$ to $270$~GHz. The samples were placed in the Faraday configuration near the end of a microwave waveguide with a rectangular cross-section of $7.0 \times 3.5$~mm$^2$ (WR~$28$) and immersed into a cryostat with a superconducting coil. The experiments were carried out at a temperature of $T =(1.5 - 4.2)$~K.  

 
\section{RESULT AND DISCUSSION} 
 
\begin{figure}[t!]
\includegraphics[width=0.47 \textwidth]{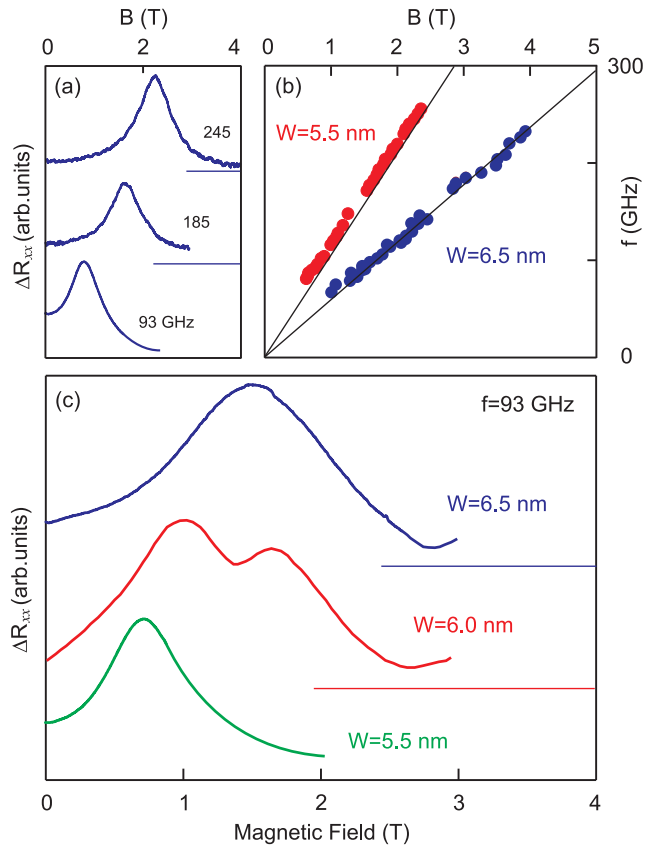}
\caption{ (a) Magnetic-field dependencies of the microwave-induced part of longitudinal resistance $\Delta R_{xx}$ at different microwave frequencies for the sample with QW width $W = 5.5$~nm. (b) Dispersions of 2D magnetoplasma excitations in AlAs QWs of widths $W=5.5$~nm (red circles) and $W=6.5$~nm (blue circles). Solid lines denote the cyclotron frequency. Lines corespond to different cyclotron masses $m_c (W = 5.5$~nm$) = 0.28 \, m_0$ and $m_c (W = 6.5$~nm$) = 0.49 \, m_0$. (c) $\Delta R_{xx} (B)$ measured for AlAs QWS of widths $W = 5.5$~nm (green curve), $W = 6.0$~nm (red curve), $W = 6.5$~nm (blue curve) at the same frequency $f = 93$~GHz.}
\label{Fig1}
\end{figure}
 
The experimental dependencies of the photoresponse $\Delta R_{xx}$ versus magnetic field $B$ obtained from a $100$ $\mu$m wide Hall bar are depicted in Fig.~\ref{Fig1}(a) for the microwave frequencies $f=93$, $185$ and $245$~GHz. The magnetic-field dependencies correspond to the sample with a QW of width $W = 5.5$~nm. The signal level without incident microwave radiation is marked by a straight line. The curves show a well-pronounced resonance that shifts with increasing frequency $f$ to higher magnetic field $B$. This suggests that the resonance corresponds to the excitation of the cyclotron magnetoplasma wave along the width of the Hall bar body~\cite{Vasiliadou:93}. Indeed, if we plot the resonance frequency $f$ as a function of the magnetic field $B$ (Fig.~1(b)), we find that the magnetoplasma resonance tends to the cyclotron resonance (CR) position $f_c=\omega_c/2 \pi= e B/2 \pi m_c$ ($m_c$ is the effective cyclotron mass of electrons).
 
Next, we present photoresponse data at $f = 93$~GHz in Fig.~\ref{Fig1}(c) measured for a series of samples with QW widths of $W = 6.5$~nm (blue trace), $W = 6.0$~nm (red trace), $W = 5.5$~nm (green trace). A drastic change in the resonance magnetic-field position and shape is observed. The samples containing QWs of $W = 5.5$~nm and $6.5$~nm exhibit a single magnetoplasma peak situated at different $B$-field values. This indicates that two different valleys are occupied by charge carriers in these samples~\cite{Dresselhaus}. This is confirmed by the photoresponse obtained for the $W = 6.0$~nm QW, where we observe two resonances at once. The latter indicates that two valleys are filled simultaneously. This is a remarkable transitional situation, when external straining or electric field can switch between $X_z$ and  $X_{x-y}$ valleys in a controllable way. The situation which lies at the heart of valleytronics.

To further elucidate the valley occupancy, we repeated the experiment of Fig.~\ref{Fig1}(a) for a series of microwave frequencies. The resultant magnetodispersions obtained for $W = 5.5$~nm (red dots) and $W = 6{.}5$~nm (blue dots) QWs are shown in Fig.~\ref{Fig1}(b). In both cases the resonance positions closely follow the CR line with a minor plasma depolarization shift. In order to precisely describe the mode magnetodispersion, we used the following expression for the magnetoplasmon frequency~\cite{Chaplik:72}:

\begin{equation}
\omega^2 = \omega_p^2 + \omega_c^2,
\label{eq.1}
\end{equation}

where $\omega_p$ is the plasmon frequency at $B = 0$~T which obeys the 2D-plasmon dispersion~\cite{Stern:67}:

\begin{equation}     
\omega_{p} = 2 \pi f_p = \sqrt{\frac{n_s e^2}{2 m_p \varepsilon_0 \varepsilon^{\ast}} q},
\label{eq.2}
\end{equation}

where $\varepsilon^{\ast}=(\varepsilon_{\rm GaAs}+1)/2$ is the effective dielectric permittivity of the surrounding medium. The plasmon wave vector $q$ for a narrow 2DES strip of width $L$ is $q= \pi N /L$ ($N=1,2, \ldots$). More accurate calculations of the zero-field plasmon frequency give $0.85 \, \omega_p$~\cite{Mikhailov:05}. Using data from Eq.~(\ref{eq.1}), we found the following values of the effective mass:  $m_c (W = 5.5$~nm$) = (0.28 \pm 0.01) \, m_0$, and  $m_c (W = 6.5$~nm$) = (0.49 \pm 0.01) \, m_0$. 

The cyclotron effective mass in anisotropic 2DES is determined as a geometric mean of the effective masses along the main crystallographic axis $m_c = \sqrt{m_{\rm l} m_{\rm tr}} = 0.47 m_0$. This value is very close to the one obtained for the QW of width $W = 6.5$~nm. Therefore, our experimental results suggest that the in-plane strongly anisotropic $X_x - X_y$ valleys are occupied for the $W = 6.5$~nm QW. In contrast, the cyclotron mass measured for the AlAs QW of width $W = 5.5$~nm is similar to the conduction band value of $m_{\rm tr} = 0.2 m_0$. Thus, the 2D electrons of this QW fill the out-of-plane isotropic $X_z$ valley. Most interestingly, we observed superposition of two resonances for the $W = 6.0$~nm QW --- one corresponding to the electrons from the $X_x - X_y$ valleys, another for the $X_z$ valley. In the $6.0$~nm AlAs QW a remarkably unique situation is realized, when both the $X_x - X_y$ valleys and the $X_z$ valley overlap energetically. A number of experiments have shown that in AlAs/AlGaAs heterostructures it is possible to lift the valley degeneracy by straining or placing the 2DES under electric or magnetic fields~\cite{Shayegan:06, Shayegan:07, Shayegan:18}. Therefore the $6.0$~nm AlAs QW is a potential platform to study new concepts in valleytronics.

\begin{figure}[t!]
\includegraphics[width=0.47 \textwidth]{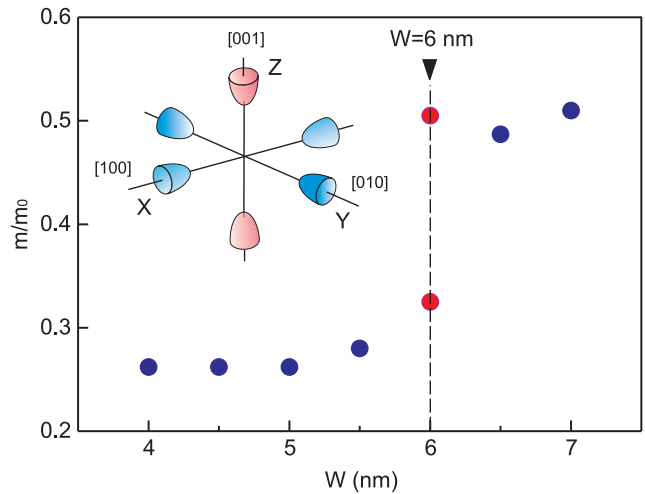}
\caption{Experimentally obtained cyclotron effective mass of charge carriers in narrow AlAs QWs as a function of their width. Dashed line marks the transition QW width $W = 6$~nm, at which the valley transition occurs. Also shown is a schematic drawing of the AlAs Fermi surface.}
\label{Fig2}
\end{figure}

Fig.~\ref{Fig2} shows a resultant dependency of 2D electrons cyclotron effective mass on the AlAs QW width. The dashed line in Fig.~\ref{Fig2} marks the transition QW width $W=6$~nm, at which the valley switching occurs. The cyclotron effective mass obtained on the AlAs QW of width $15$~nm equaled $m_c (W = 15$~nm$) = (0.47 \pm 0.01) \, m_0$ and is not shown in the Fig.~\ref{Fig2}. Notably, the obtained value of the  $X_z$ effective mass $m_c (W = 5.5$~nm$) = (0.28 \pm 0.01) \, m_0$ is well above the value reported in experiments performed on wide AlAs QWs of widths $11$~nm and $15$~nm~\cite{Gunawan:04, Muravev:15}. Also $m_c (W = 5.5$~nm$)$ exceeds AlAs volume band effective mass of $m^{\ast}=0.2 \, m_0$~\cite{Adachi:85}. The mass of charge carriers is one of the main parameters of electronic devices. Thus it is instructive to understand the physical origin of the observed phenomenon. Previous studies ruled out non-parabolicity and retardation effects as possible explanations for the mass enhancement~\cite{Yamada:94, Stadt:96, Shayegan:04, Shayegan:09, Khisameeva:18}. Our present experimental data shown in Fig.~\ref{Fig1}(c) reveals that the cyclotron mass is almost independent of the QW width $W$. This finding excludes the influence of AlGaAs barrier and electron-electron interaction on the magnitude of the 2D electrons effective mass in the QW. 

\begin{figure}[t!]
\includegraphics[width=0.47 \textwidth]{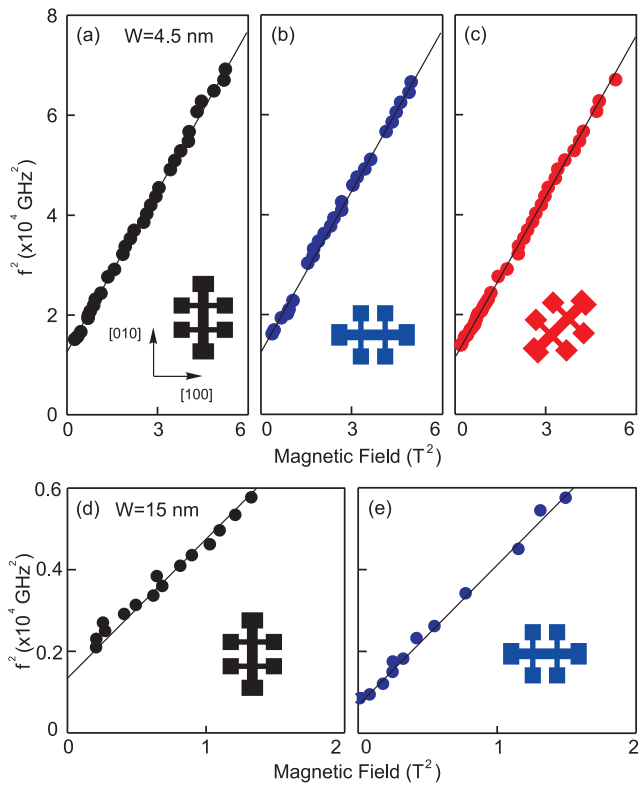}
\caption{Magnetodispersions $f^2 (B^2)$ of plasma excitations in AlAs QW of width $W = 4.5$~nm in Hall bars with $L=20$~$\mu$m oriented along $[010]$ (black circles), $[100]$ (blue circles), and $[110]$ (red circles). $f^2 (B^2)$ in AlAs QW of width $W = 15$~nm in Hall bars with $L=50$~$\mu$m oriented along $[010]$ (black circles), $[100]$ (blue circles).
The scales of graphs (a)-(c) and (d)-(e) are identical. The solid line represents the theoretical prediction according to Eq.~(\ref{eq.1}). The configuration of Hall bars are schematically presented as an inset.}
\label{Fig3}
\end{figure}  

To further prove that 2D electrons in narrow AlAs QWs of width $W<6$~nm occupy only a single $X_z$ valley, we investigated isotropism of their Fermi contour. Indeed, the $X_z$ valley possesses in-plane isotropic Fermi contour. Therefore, if no mixing with highly in-plane anisotropic $X_x - X_y$ valleys takes place, we should measure the isotropic plasmon effective mass $m_p$. To do this, we fabricated three Hall bar structures ($L = 20$~$\mu$m) from the $W = 4.5$~nm QW along different crystallographic directions (Fig.~\ref{Fig3}). Notably, the cyclotron effective mass $m_c=1/2 \pi (\partial S_p/\partial E)$ is not a good indicator of the Fermi contour isotropism, because it is determined by the whole cross section of the Fermi contour. Fig.~\ref{Fig3} displays $f^2 (B^2)$ dependence of the cyclotron magnetoplasma mode excited in the Hall bars oriented along the $[010]$ (a), $[100]$ (b), and $[110]$ (c) crystallographic axes. The zero-field plasmon frequency $f_p$ carries unique information about the charge effective mass along the direction across the Hall bar body. In order to obtain the zero-field frequencies $f_p$ the experimental data of Fig.~\ref{Fig3}(a-c)  were fitted using Eq.~(\ref{eq.1}) (the solid line in Fig.~\ref{Fig3}). From this fitting we found that $f_p=\omega_p/2 \pi = 110$~GHz ($[100]$), $f_p=\omega_p/2 \pi = 111$~GHz ($[010]$), and $f_p=\omega_p/2 \pi = 108$~GHz ($[\bar{1}10]$). Corresponding effective masses were $m_p ([100]) = (0.27 \pm 0.01) \, m_0$, $m_p ([010]) = (0.26 \pm 0.01) \, m_0$, and $m_p ([\bar{1}10]) = (0.27 \pm 0.01) \, m_0$. Obtained values of $m_p$ confirm that charge carriers in narrow AlAs QWs ($W < 6$~nm) reside in a single isotropic $X_z$ valley. 

To further illustrate the proposed method to study Fermi contour isotropism, we conducted the same experiment with Hall bars ($L = 50$~$\mu$m) fabricated from the wide $W = 15$~nm AlAs quantum well. Corresponding data measured for Hall bars oriented along the $[010]$ and $[100]$ crystallographic directions is shown in Fig.~\ref{Fig3}(d-e). Unlike the $W = 4.5$~nm QW case the zero-field plasmon frequencies in Fig.~\ref{Fig3}(d-e) differ significantly  $f_p = 38$~GHz ($[100]$) and   $f_p = 26$~GHz ($[010]$). This finding indicated anisotropy of occupied $X_x - X_y$ valleys. However, a detailed analysis of the data is a bit challenging due to the energy splitting between $X_x$ and $X_y$ valleys. Indeed, the residual in-plane strain lifts the $X_x$ and $X_y$ valley degeneracy, leading to an intervalley energy splitting~\cite{Shayegan:06}. The total density $n_s=n_x+n_y$ is then distributed between both $X_x$ and $X_y$ valleys. Obtained plasmon spectra allow to unequivocally determine $n_x$ and $n_y$, after which it becomes possible to determine $m_{\rm tr}$ and $m_{\rm l}$ (for a detailed procedure we refer to~\cite{Muravev:15}). Finally, we get $m_{\rm tr}= (0.22 \pm 0.04) \, m_0$ and $m_{\rm l}= (1.2 \pm 0.2) \, m_0$.

\section{CONCLUSION} 
 
In summary, we investigated resonant microwave absorption in Hall bar structures containing 2D electrons in AlAs QWs of different widths. In total, eight structures with different QW widths from $4.0$~nm to $15$~nm have been studied. We observed $\Gamma - X$ energy spectrum transition and identified an exact value of the QW width ($W = 6$~nm) at which valley balance is achieved. Our experiments unambiguously eliminated the influence of AlGaAs barrier band structure on the 2D electrons effective mass in the QW. Moreover, we experimentally proved an isotropic Fermi contour of the $X_z$ valley in the QW plane. These results pave the way for a new concepts of valleytronic devices.

\section{ACKNOWLEDGMENTS}

The scientific result of the present manuscript has been obtained within the framework of the Russian Science Foundation (Grant No.~18-72-10072).



\end{document}